\documentclass[prd,aps,preprint,amsmath,nofootinbib,amssymb,eqsecnum,showkeys,tightenlines]{revtex4-1}
\pdfoutput=1
\usepackage{verbatim,graphics,graphicx,color,slashed}
\usepackage{ulem} % this one is for cross out text
\usepackage[colorlinks=true,
            linkcolor=red,
            urlcolor=blue,
            citecolor=blue]{hyperref}

\graphicspath{{figures/}}

\newcommand{\GeV}{\mathrm{GeV}}
\newcommand{\TeV}{\mathrm{TeV}}
\newcommand{\PeV}{\mathrm{PeV}}

\begin{document}
%\title{Heavy Neutrino Mediated Dark Matter Decays and IceCube Events}
\title{IceCube Events from Heavy DM decays\\ through the Right-handed Neutrino Portal}
\author{P. Ko, Yong Tang}
\affiliation{School of Physics, Korea Institute for Advanced Study, \\
 Seoul 130-722, Korea}
\date{\today}

\begin{abstract}  
The recently observed IceCube PeV events could be due to heavy dark matter (DM) decay. 
In this paper, we propose a simple DM model with extra $U(1)_X$ gauge symmetry and 
bridge it with standard model particles through heavy right-handed neutrino. 
The Dirac fermion DM $\chi$ with mass $\sim 5$ PeV can dominantly decay into a dark 
Higgs ($\phi$), the SM Higgs ($h$) and a neutrino ($\nu$).  If the lifetime of $\chi$ is $\sim O(10^{28})$ sec, the resulting neutrino flux can fit data consistently.  
The neutrino flux from $\chi \rightarrow \phi h \nu$ in our model is softer than the one predicted from $\chi \rightarrow \nu h$, for example.
We also discuss a possible mechanism to produce DM with the right relic abundance. 
\end{abstract}

\maketitle

%----------------------------------------------------------------------------------
\section{Introduction}\label{sec:intro}
%----------------------------------------------------------------------------------
Recently, the IceCube Collaboration has reported the detection of 37 neutrino events with 
energy between $30 ~\TeV - 2 ~\PeV$, of which three have energy above 1 PeV 
~\cite{IceCube:2014gkd,IceCube:2013bka,IceCube:2013jdh}. 
According to the recent analyses~\cite{Aartsen:2015ivb, Palomares-Ruiz:2015mka, Bustamante:2015waa,Huang:2015flc}, the three-year 
data are consistent with equal fluxes of all three neutrino flavors and with isotropic arrival  directions. However, the neutrino flux required to fit the data in $100\TeV - \PeV$ range is  around $10^{-8}~\GeV$ cm$^{-2}$s$^{-1}$sr$^{-1}$ per flavor, and rejects a purely 
atmospheric explanation at $5.7\sigma$.  Therefore astrophysical and/or new physics 
explanations have been pursued for the origin of these high energy neutrinos.

Possible astrophysical sources are involved with supernova remnants (SNR)~\cite{Chakraborty:2015sta, Loeb:2006tw,Murase:2013rfa, Chang:2014sua, Liu:2013wia}, active galactic 
nuclei (AGN)
~\cite{Kalashev:2013vba, Kalashev:2014vya, Stecker:1991vm, Essey:2009ju},  
and gamma-ray bursts (GRB)
~\cite{Waxman:1997ti, Murase:2013ffa}, all of which assume some specific emission spectra due to different 
production environments. In a model independent analysis, 
the best-fit power law spectrum from the IceCube analysis~\cite{IceCube:2014gkd} is 
$E^2_\nu d\Phi_\nu/dE_\nu \simeq 1.5\times 10^{-8}(E_\nu/100~\TeV)^{-0.3}$ cm$^{-2}$
s$^{-1}$sr$^{-1}$. It is not very straightforward to fit such a spectrum  by astrophysical 
sources. In all cases, extragalactic sources are needed due to the isotropic feature and 
galactic constraints~\cite{Ahlers:2013xia, Razzaque:2013uoa, Wang:2014jca, Ahlers:2015moa}. Then identifying such astrophysical sources will be crucial for further understanding of IceCube events. 

Dark matter (DM) and other new physics interpretations have been also investigated in 
various ways or models. Heavy DM might decay into SM particles that give energetic PeV 
neutrinos~\cite{Yanagida:2013kka, Esmaili:2013gha, Ema:2013nda, Esmaili:2014rma, Ema:2014ufa, Fong:2014bsa, Higaki:2014dwa, Bhattacharya:2014vwa, Anchordoqui:2015lqa, Boucenna:2015tra, Aisati:2015vma, Roland:2015yoa, Rott:2014kfa, Murase:2015gea, Berezhiani:2015fba}~\footnote{If involving of dark halo substructure, Ref.~\cite{Zavala:2014dla} showed annihilating DM scenario may also be possible.}, or it could decay into some light DM particles which interact with nucleons and mimic neutrino events~\cite{Bhattacharya:2014yha, Kopp:2015bfa}. The resulting neutrino flux would
still be consistent with isotropy so far, since galactic and extragalactic DM contribute at the similar order. One unique feature of DM explanation is that there should 
be sharp energy cut-off in the neutrino spectrum, which could be tested by future data. 
Also, a possible gap around 400 TeV $\sim$ 1 PeV, although not statistically significant yet, motivated considerations of new interactions, two-component flux and leptophilic DM decay~\cite{Barger:2013pla, He:2013zpa, Ioka:2014kca, Araki:2014ona, Cherry:2014xra, Chen:2014gxa, Kamada:2015era, Boucenna:2015tra}. 

In this paper, we propose a simple DM model to explain the IceCube PeV events. 
A dark sector with new $U(1)_X$ gauge symmetry is introduced and can have connection 
with the standard model (SM) sector through neutrino-portal interactions as well as the 
Higgs portal interaction.  Fermionic DM ($\chi$) has $\sim$ PeV mass and mostly decays 
into three-body final state with dark Higgs ($\phi$), SM Higgs ($h$) and active neutrino. 
The produced neutrinos from primary $\chi$ decay and the secondary $h$ and $\phi$ 
decays can explain the observed PeV event spectra, while the atmospheric and astrophysical neutrinos are included for the low-energy part. 

This paper is organized as follows. In Section.~\ref{sec:model} we introduce our DM model with dark $U(1)_X$ gauge symmetry, heavy right-handed neutrino portal and Higgs portal 
interactions. In Section.~\ref{sec:spectrum} we outline the general formalism for calculating neutrino flux from galactic and extragalactic DM decay. In Section.~\ref{sec:numerical} we present both total and differential decay width for the relevant three-body decay in our model and compare the numerical results with IceCube data. In Section.~\ref{sec:relic} we discuss 
a possible mechanism to generate the correct relic density for DM and direct/indirect 
detection constraints. Finally, we give our conclusions.

%----------------------------------------------------------------------------------
\section{Model}\label{sec:model}
%----------------------------------------------------------------------------------

We consider a dark sector with a dark Higgs field $\Phi$ and a Dirac fermion DM $\chi$ 
associated $U(1)_X$ gauge symmetry. Their $U(1)_X$ charges are assigned as follows~\footnote{A similar setup with different dark charge assignments has been considered 
for the AMS02 positron excess~\cite{Ko:2014lsa}. One may also use discrete symmetries, see Ref.~\cite{Kang:2010ha} for example.}:
\[
(Q_\Phi, Q_\chi) = (1,1) .
\]
We begin with the following renormalizable and gauge invariant  Lagrangian including 
just one singlet right-handed (RH) neutrino $N$ and one lepton flavor (more $N$s 
and/or flavors can be easily generalized): 
%\begin{widetext} 
\begin{align}
\mathcal{L}=& \mathcal{L}_{\mathrm{SM}} + \frac{1}{2}\bar{N}i\slashed{\partial}N - 
\left(\frac{1}{2}m_{N} \bar{N}^c N + y \bar{L}\widetilde{H} N  + \textrm{h.c.}\right) 
%\nonumber \\
-\frac{1}{4}X_{\mu\nu}X^{\mu\nu} -\frac{1}{2}\sin{\epsilon}X_{\mu\nu}F^{\mu\nu}_{Y}  
\nonumber\\ & + D_\mu \Phi^{\dagger}D^\mu \Phi -V(\Phi,H)
+ \bar{\chi}\left(i\slashed{D} - m_\chi\right)\chi  - \left(f \bar{\chi}\Phi N  +\textrm{h.c.}\right), 
\end{align}
%\end{widetext}
where $L = (\nu \; l)^T$ is a left-handed (LH) SM $SU(2)$ lepton doublet, 
$H$ is the SM Higgs doublet,   
$X_{\mu\nu}=\partial_\mu X_\nu - \partial_\nu X_\mu$ is the field strength for $U(1)_X$ 
gauge field $X_\mu$, $F^{\mu\nu}_Y$ is for SM hypercharge $U(1)_Y$, and $\epsilon$ 
is the kinetic mixing parameter.   Two types of Yukawa couplings, $y$ and $f$, can be taken 
as real parameters, ignoring CP violation for simplicity. We define covariant derivative as 
$ D_\mu  = \partial_\mu - i g_X X_\mu $. Since we are interested in explaining the IceCube 
PeV events in terms of DM $\chi$ decay, we shall take $m_\chi \sim \PeV$. 
Other parameters in our model are free variables. 

The scalar potential $V$ of this model is given by %parametrized as
%\begin{widetext}
\begin{equation}
V=\lambda_H \left(H^\dagger H -\frac{v^2_H}{2}\right)^2+\lambda_{\phi H}\left(H^\dagger H -\frac{v^2_H}{2}\right) \left(\Phi^\dagger \Phi -\frac{v^2_\phi}{2}\right) + \lambda_\phi \left(\Phi^\dagger \Phi -\frac{v^2_\phi}{2}\right)^2,
\end{equation}
%\end{widetext}
Both electroweak and dark gauge symmetries are spontaneously broken by the nonzero 
vacuum expectations values of $H$ and $\Phi$: %state, 
$\langle H\rangle=\left( 0,\; v_{H}/\sqrt{2}\right)^{T},\;\langle\Phi\rangle= v_{\phi}/\sqrt{2} \ .$
Here $v_H\simeq 246$GeV is the same as SM value but $v_{\phi}$ might be taken 
as a free parameter. In the unitarity gauge, we can replace the scalar fields with
\begin{equation}
 H\rightarrow \frac{1}{\sqrt{2}}
 \left(
	 \begin{array}{c}
		 0\\ v_{H}+h(x) 
	 \end{array}
 \right)
  ~~ {\rm and}~~
 \Phi\rightarrow\dfrac{v_\phi+\phi(x)}{\sqrt{2}} .
\end{equation}
Note that $h$ and $\phi$ shall mix with each other thanks to the Higgs-portal operator, 
(the $\lambda_{\phi H}$ term) \footnote{The $\lambda_{h\phi}$ term can also help to stabilize the 
electroweak vacuum~\cite{Tang:2013bz,Chen:2012faa,Baek:2012uj,Baek:2012se}.}. Through this mixing, $\phi$ can decay into SM particles. Another important mixing happens among three neutral gauge bosons, photon $A_\mu$, $Z_\mu$ and $X_\mu$. Such a mixture would enable an extra mass eigenstate $Z_{\mu}'$ (mostly $X_\mu$) to decay SM fermion pairs.  Then DM $\chi$ 
scattering off nucleus is possible by the $Z'$ exchange, and the cross section essentially 
depends on $\epsilon,v_\phi, m_{Z'}$. It is easy to choose small $\epsilon$, or heavy 
masses to evade the constraints from DM direct detection~\cite{Baek:2014goa}.

When the right-handed neutrino $N$ is much heavier than $\chi$, we can integrate it out
and obtain an effective operator, 
\begin{equation}
\frac{yf}{m_N}\bar{\chi}\Phi H^\dagger L + h.c.,
\end{equation}
which would make $\chi$ decay possible but long lived. After spontaneous gauge symmetry breaking, we have several higher dimensional effective operators from the aforementioned 
operator Eq. (2.4) as follows: 
\begin{align}\label{eq:operator}
\frac{v_\phi v_H}{m_N}\bar{\chi}\nu,\;\frac{v_\phi}{m_N}\bar{\chi}h\nu ,\;
\frac{v_H}{m_N}\bar{\chi}\phi \nu,\;&\frac{1}{ m_N}\bar{\chi}\phi h\nu,
\end{align}
with the common factor $\dfrac{yf}{2}$ for all these operators. If kinematically allowed, all the 
above operators induce $\chi$ decays into different channels with fixed relative branching 
ratios.  Within the heavy $\chi$ limit, $m_\chi\gg m_\phi, m_{Z'}, m_h, m_Z, m_W$, 
the mass operator  $\bar{\chi}\nu$ in Eq.~(\ref{eq:operator}) would induce a tiny 
mixing between $\chi$ and $\nu$ with the mixing angle $\beta$ approximately given by
\begin{equation}
\beta \simeq \frac{yf}{2}\frac{v_\phi v_H}{m_N m_\chi} \ . 
\end{equation}
Then the gauge interactions for $\chi$ and $\nu$ will generate the decay channels,
\begin{equation}
\chi\rightarrow Z'\nu, Z\nu, W^{\mp}l^{\pm},
\end{equation}
with their branching ratios being proportional to  $\sim v^2_H:v^2_\phi:2v^2_\phi$. Two dim-4 operators, $\bar{\chi}h\nu$ and $\bar{\chi}\phi \nu$, would lead $\chi$ to the following decays,
\begin{equation}
\chi\rightarrow h\nu,\phi \nu,
\end{equation}
with their branching ratios being proportional to $\sim v^2_\phi : v^2_H$. 
It is also straightforward to get the following relation for the branching ratios, 
\begin{equation}
Br(\chi \rightarrow \phi \nu):Br(\chi \rightarrow Z' \nu) \simeq 1:1 \ .
\end{equation}
Therefore,  all the decay branching ratios are basically calculable and completely 
fixed in this model \footnote{This is also true in the model for the AMS02 positron excess~
\cite{Ko:2014lsa}.}.  Note that the decay modes with $Z'$ or $\phi$ are unique features of   
DM models with dark gauge symmetries \footnote{This is also true of three-body decays 
of DM discussed in the following paragraph.}.

Another interesting phenomenon in this model is that three body decay channel 
$\chi\rightarrow \phi h \nu$ is dominant over all other channels when $m_\chi\gg v_\phi$:
\begin{equation}
\frac{\Gamma_3\left(\chi\rightarrow \phi h \nu \right)}{\Gamma_2\left(\chi\rightarrow h \nu ,\phi \nu \right)}\simeq \frac{1}{16\pi^2} \frac{m^2_\chi}{v^2_\phi+v^2_H} \gg 1,
\end{equation}
since we actually have an enhancement from heavy $m_\chi$ even though there is a phase space suppression from three-body final states.   
There are another three-body decay channels that are equally important: 
\begin{equation*}
\chi \rightarrow \phi/Z' + h +\nu,\; \phi/Z' + Z + \nu,\; \phi/Z' + W^{\pm} +l^{\mp},
\end{equation*}
with branching ratios $1:1:2$ due to the Goldstone boson equivalence theorem. 
In the following, if not otherwise stated explicitly, we use $\chi\rightarrow \phi h \nu$ to represent all these 
channels and in numerical calculations we take all of them into account.

To give a rough impression for the relevant 
parameter ranges, we can perform an order-of-magnitude estimation (complete formulas 
and details of calculation are given in the Appendix): 
\begin{eqnarray}\label{eq:yfnrelation}
\Gamma_3\left(\chi\rightarrow \phi h \nu \right) & \sim & 
\frac{m_\chi^3}{96 \pi^3}\left(\frac{yf}{m_N}\right)^2 \sim \frac{1}{10^{28} \textrm{sec}}
\\ & \Rightarrow & \frac{yf}{m_N} \sim 10^{-36} \GeV^{-1},
\end{eqnarray}
which shows the required value for the combination $yf/m_N$. 
If one additionally assumes that $N$ should be responsible for neutrino mass through the 
usual Type-I see-saw mechanism, then we have $y \sim 10^{-5} \sqrt{m_N/\PeV}$. 
Therefore we would have $y\sim 1$ and $f\sim 10^{-22}$ for $m_N\sim 10^{14}\GeV$ and, 
$y\sim 10^{-5}$ and $f\sim 10^{-25}$ for $m_N\sim \PeV$. In any case, $f$ is very tiny and 
seems unnaturally small. However, it is still natural a la 't Hooft~\cite{tHooft:naturalness} since taking $f=0$ 
enhances symmetries of the theory, namely DM number conservation 
\footnote{The DM current $j_\chi^\mu = \bar{\chi} \gamma^\mu \chi$ is conserved 
in the limit $f \rightarrow 0$.}.  
In later discussion, we shall take $y,f,m_N$ as free parameters, unless specified. 

%----------------------------------------------------------------------------------
\section{Neutrino Flux from DM decay}\label{sec:spectrum}
%----------------------------------------------------------------------------------
The neutrino flux from dark matter decay is composed of galactic and extragalactic contributions which are equally important as we shall see below. Galactic neutrino flux 
at kinetic energy $E$ from DM decay in our Milky Way dark halo is given by
\begin{equation}\label{eq:flux_g}
\left.\frac{d\Phi_\nu^G}{dE_\nu}\right|_{E_\nu=E}=\frac{1}{4\pi}\sum_{i}\Gamma_i\int_0^\infty 
dr  \frac{\rho^G_\chi \left(r'\right)}{m_\chi} \left.\frac{dN^i_\nu}{d E_\nu} \right|_{E_\nu=E},
\end{equation}
where $\Gamma_i$ is partial width for decay channel $i$, $d N^i_\nu/dE_\nu$ is the neutrino spectrum at production, $r'=\sqrt{r_\odot^2 + r^2 -2r_\odot  r \cos \theta}$, $r$ is the distance to earth from the DM decay point,  $r_\odot\simeq 8.5$kpc for the solar system and $\theta$ is the observation angle between the line-of-sight and the center of the Milky Way. For the galactic DM density distribution, we use the following standard NFW profile~\cite{Navarro:1995iw},
\begin{equation}\label{eq:NFW}
\rho^G_\chi\left(r'\right)=\rho_\odot \left[\frac{r_\odot}{r'}\right] \left[\frac{1+r_\odot/r_c}{1+r'/r_c}\right]^{2},
\end{equation}
with parameters $r_c\simeq 20$ kpc and $\rho_\odot \simeq 0.4~\textrm{GeV}/\textrm{cm}^3$. For decaying dark matter the flux is not very sensitive to DM density profile, so our discussions and results will still apply if another different profile is used. 

We can also get the extragalactic or cosmic contribution from a formula similar to the above one, by taking cosmic expansion into account, namely the red-shift effect~\cite{Esmaili:2013gha}:
\begin{equation}\label{eq:flux_eg}
\left.\frac{d\Phi_\nu^{EG}}{dE_\nu}\right|_{E_\nu=E}=\frac{\rho_c\Omega_{\chi}}{4\pi m_\chi}\sum_{i}\Gamma_i\int_0^\infty \frac{dz}{\mathcal{H}} \left.\frac{dN^i_\nu}{d E_\nu} \right|_{E_\nu=(1+z)E},
\end{equation}
where $E'$ is red-shifted to $E$ as $E'=(1+z)E$, the red-shift $z$ is defined as $1+z=a_0/a$ with present scale factor $a_0$ being normalized to $1$, the critical energy density $\rho_c=5.5\times 10^{-6} \GeV/\mathrm{cm}^{3}$ and $\Omega_{\chi}\simeq 0.27$ is DM $\chi$'s fraction. The Hubble parameter $\mathcal{H}$ is related to its present value through
\[
 \mathcal{H}=\mathcal{H}_0\sqrt{\Omega_\Lambda + \Omega_\mathrm{m}(1+z)^3 + \Omega_{r}(1+z)^4},
\]
$\Omega_\lambda$, $\Omega_\mathrm{m}$ and $\Omega_{r}$ are energy fractions of dark energy, all matter, and radiations, respectively. We shall use the latest results from \texttt{Planck}~\cite{Planck:2015xua} for numerical evaluation. 

%----------------------------------------------------------------------------------
\section{Numerical Results}\label{sec:numerical}
%----------------------------------------------------------------------------------

\begin{figure}[t]
\includegraphics[width=0.40\textwidth, height=0.41\textwidth]{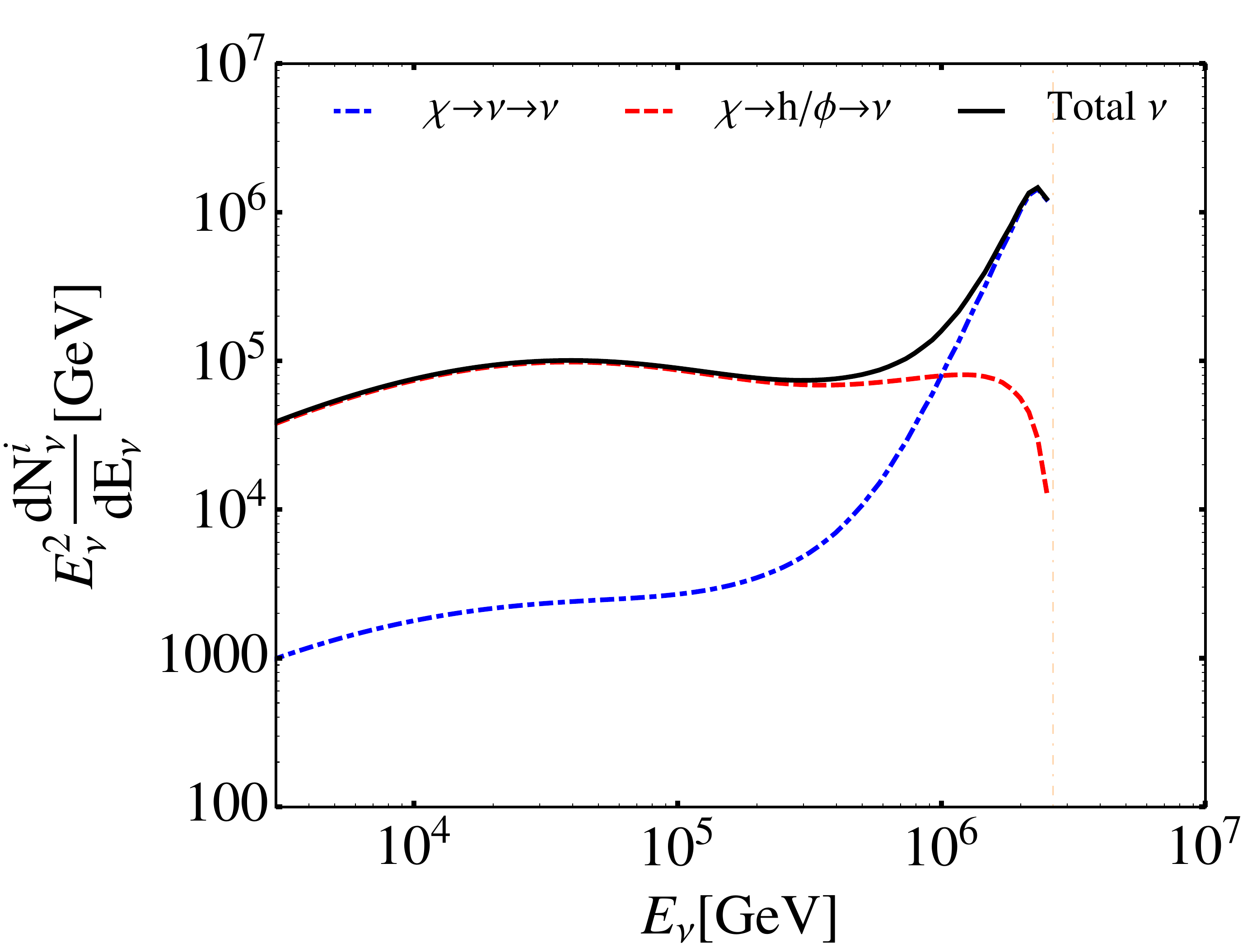} 
\includegraphics[width=0.40\textwidth, height=0.41\textwidth]{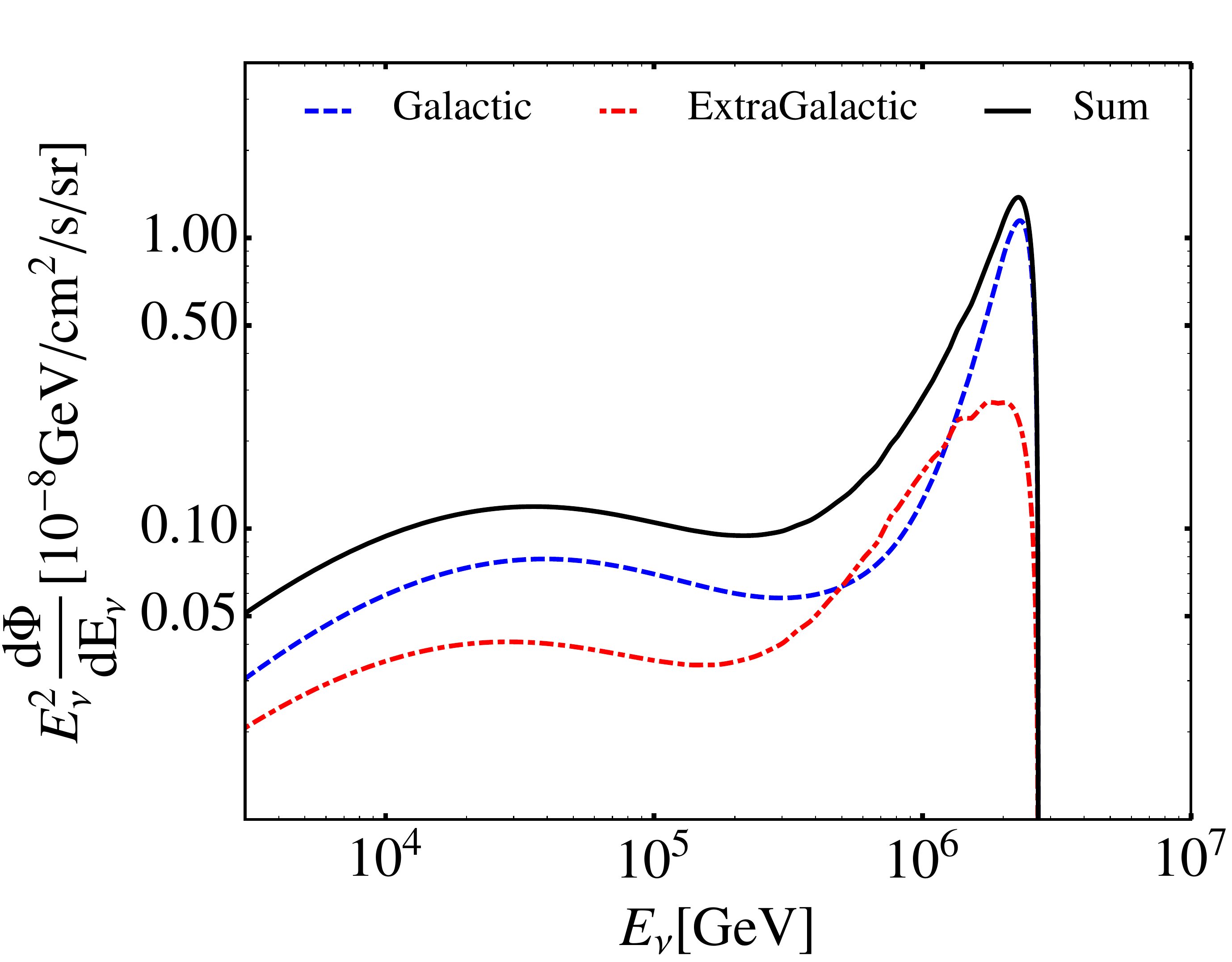}
\caption{Neutrino spectra from DM $\chi$ decay with $m_\chi\sim 5~\PeV$ and lifetime 
$\tau_\chi =1/\Gamma \sim 2\times 10^{28}$s. The left panel shows individual contribution 
of different final states from $\chi$'s decay, $\nu$ (blue dot-dashed curve) and $h/\phi$ 
(red dashed curve), respectively. The right panel presents the galactic (blue dashed curve) 
and extragalactic (red dot-dashed curve) neutrino flux. 
\label{fig:spectrum}}
\end{figure}

To compute the neutrino flux from DM decay, we first need to calculate the total and 
differential three-body decay width for $\chi \rightarrow \phi + h + \nu$. 
In the heavy $\chi$ limit, we have obtained the total width
\begin{align}
\Gamma \simeq  \frac{m_\chi^3}{768 \pi^3}\left(\frac{yf}{m_N}\right)^2, 
\end{align}
and normalized differential decay widths 
\begin{align}
\frac{1}{\Gamma}\frac{d\Gamma}{dE_\nu} & \simeq 24 E_\nu^2/m_\chi^3, \; 0<E_\nu < m_\chi/2, \\
\frac{1}{\Gamma}\frac{d\Gamma}{dE_h}   & \simeq 12 E_h \left(m_\chi - E_h \right)/m_\chi^3,\; 0<E_h < m_\chi/2, \\
\frac{1}{\Gamma}\frac{d\Gamma}{dE_\phi}& \simeq 12 E_\phi \left(m_\chi - E_\phi \right)/m_\chi^3,\; 0<E_\phi < m_\chi/2.
\end{align}
The details of the calculation are given in the Appendix where complete formulas with nonzero mass parameters are also presented. The above differential widths are essential ingredients to get the final neutrino flux. For example, $\nu$s from different decay final states are given by
\begin{equation}
\frac{d N}{dE}\left({x \rightarrow \nu }\right)= \int \frac{1}{\Gamma}\frac{d\Gamma}{dE_{x}} \frac{d N_\nu\left(E_x \right)}{dE} d E_x,
\end{equation}
where  $x=\nu,h,W,Z,Z',\phi$. Note that  $d N_\nu\left(E_x \right)/dE $ in the integrand can be calculated with \texttt{Pythia}~\cite{Pythia64:2006za} or \texttt{PPPC4DMID}~\cite{PPPC4DMID:2010xx}.

In Fig.~\ref{fig:spectrum} we show the neutrino spectra. Just for illustration, we choose the mass of DM $\chi$ around $5$ PeV and its lifetime $2\times 10^{28}$s. The neutrino spectra are multiplied by $E^2_\nu$ in order to account for the energy dependence of the neutrino-nucleus cross section  so that it might be compared with 
IceCube data more easily. 
The blue dot-dashed curve indicates the spectrum from final $\nu$ in $\chi$'s decay and 
red dashed one marks $h/\phi$'s contribution. For simplicity, the mass of $\phi$ has been chosen to be just as the SM Higgs mass. However, other choice does not affect our result much, since  in the left panel of Fig.~\ref{fig:spectrum} we see that in the high energy part 
$h/\phi$'s contributions are basically negligible and it is the high energy part that explains 
the IceCube PeV events. In the right panel, we include the red-shifted effects and show 
both galactic (blue dashed) and extragalactic (red dot-dashed) contributions. 

Next, we compare our model predictions with \texttt{IceCube} three-year data
~\cite{IceCube:2014gkd}. To parameterize the possible astrophysical neutrino
fluxes at low energy, we consider either a broken power law (BPL) or unbroken power law 
(UPL)~\cite{Boucenna:2015tra}, 
\begin{eqnarray}
E^2_\nu \frac{d\Phi_{\textrm{bkg}}}{dE_\nu}&=& J_0^\mathrm{{BPL}}\left(\frac{E_\nu}{100 \TeV}\right)^{-\gamma_1}\mathrm{exp}\left(-\frac{E_\nu}{E_0}\right),\\
E^2_\nu \frac{d\Phi_{\textrm{bkg}}}{dE_\nu}&=& J_0^\mathrm{{UPL}}\left(\frac{E_\nu}{100 \TeV}\right)^{-\gamma_2},
\end{eqnarray}
where the first one has an exponential cut-off at energy scale $E_0$ which is chosen to be $125$ TeV in agreement with the SNR results~\cite{Chakraborty:2015sta}.

\begin{figure}[t]
\includegraphics[width=0.40\textwidth, height=0.41\textwidth]{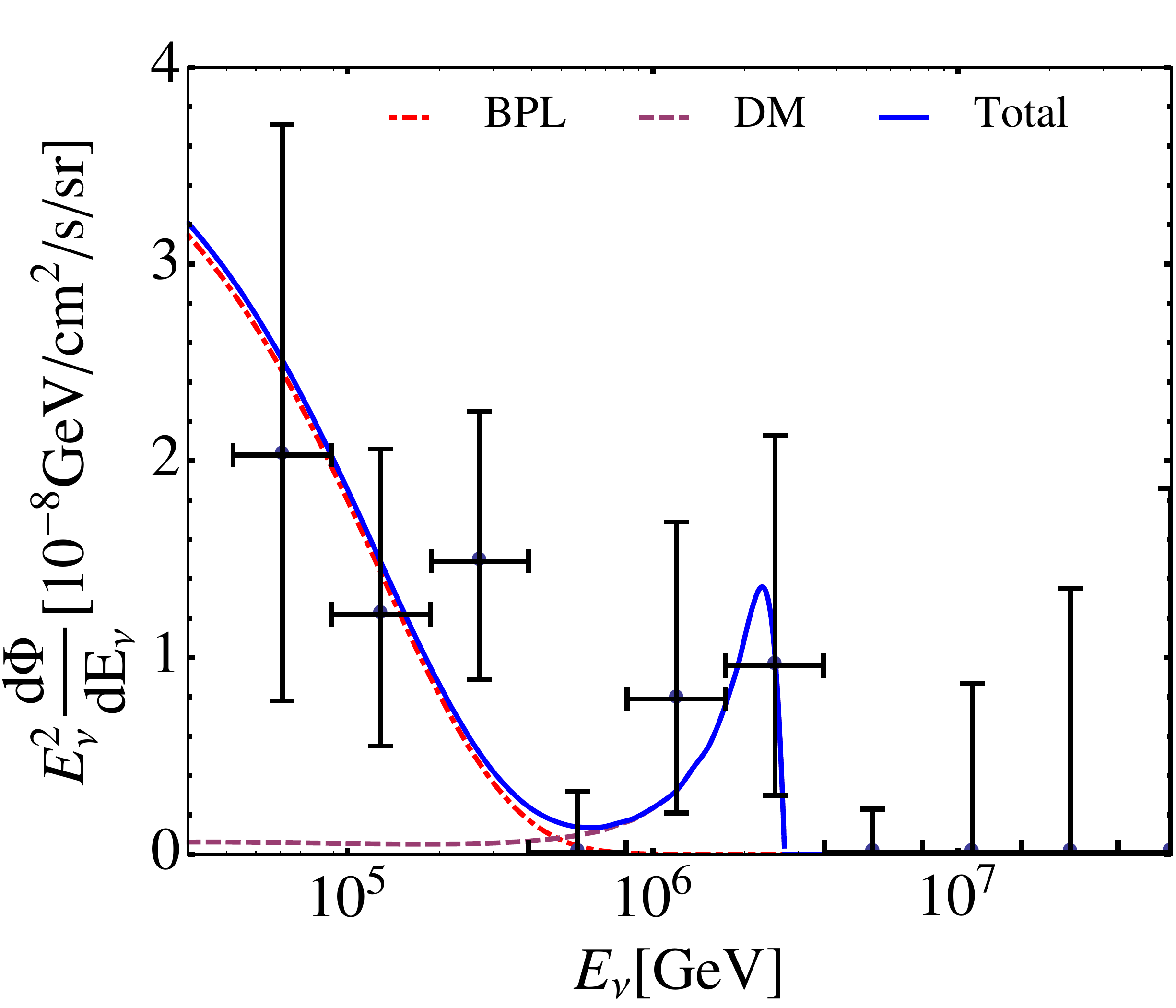} 
\includegraphics[width=0.40\textwidth, height=0.41\textwidth]{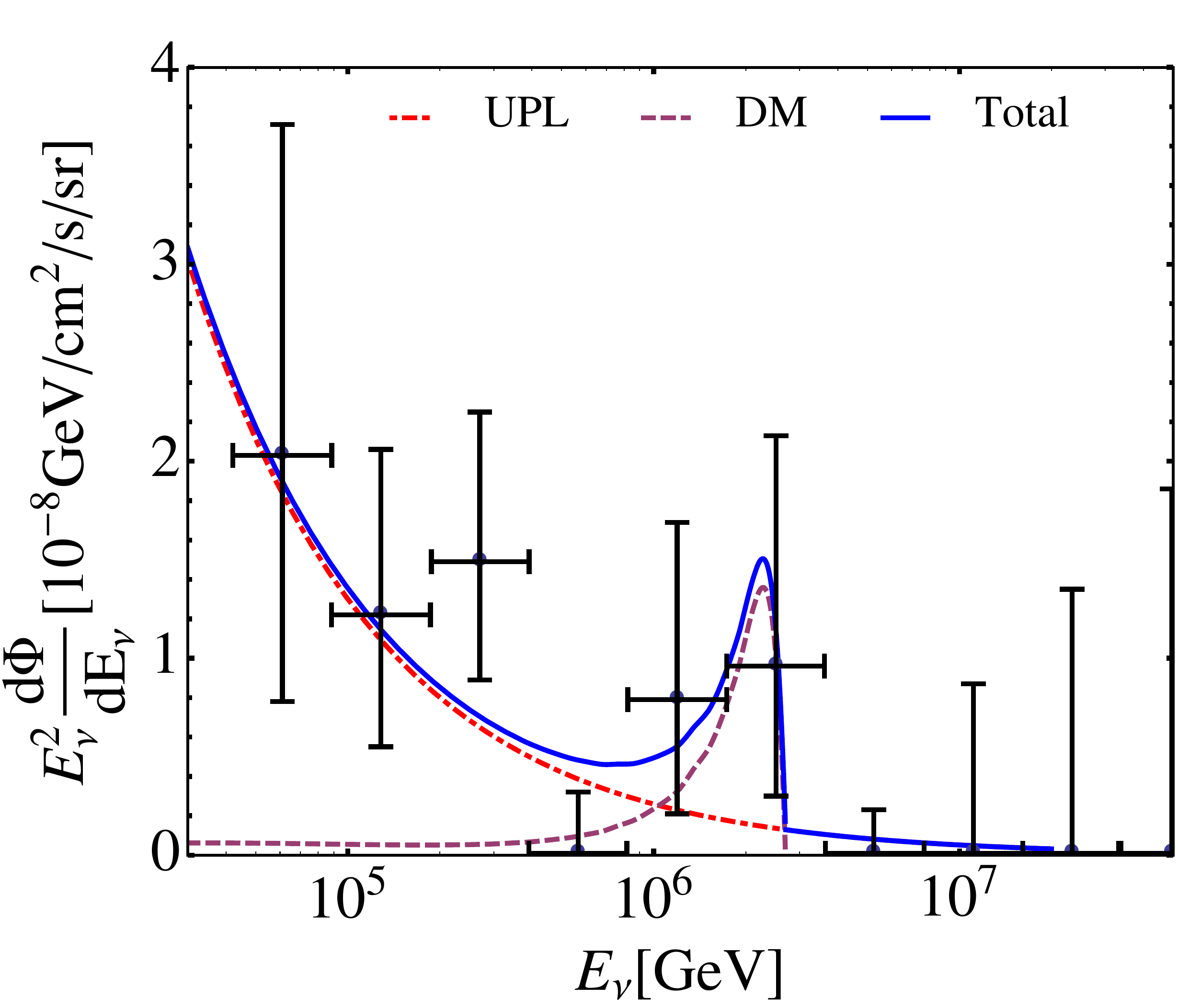}
\caption{Neutrino flux from DM $\chi$'s decay with $m_\chi\sim 5\PeV$ and lifetime $\tau_\chi =1/\Gamma \sim 2\times 10^{28}$s and IceCube Data~\cite{IceCube:2014gkd}. The left (right) panel used a broken (unbroken) power law (BPL) for astrophysical neutrino flux with a red dot-dashed curve. DM's contributions and total flux are labeled with purple dashed and blue solid curves, respectively. See details in the text.  \label{fig:flux}}
\end{figure}

We illustrate in Fig.~\ref{fig:flux} with two different astrophysical flux choices. The low energy data are best fitted by varying $J_0$ and the spectral index $\gamma$, but the high energy PeV data points are fit by DM decay. 
In the left panel, we have used $J_0^\mathrm{{BPL}}=4.1\times 10^{-8}\GeV$/cm$^2$/s/sr and $\gamma_1=0$(red dot-dashed curve). In the right panel, we have used $J_0^\mathrm{{UPL}}=1.3\times 10^{-8}\GeV$/cm$^2$/s/sr and $\gamma_2=0.7$(red dot-dashed curve). DM's contributions and total flux are labeled with purple dashed and blue solid curves, respectively. As we can see in the figure, our model can agree with the PeV data. Also a gap could appear around 400 TeV although the current data can not tell existence of the gap is statistically significant. The feature for the DM decay spectrum is that there would be a  sudden drop around $E_\nu \simeq m_\chi/2$. 
As more data are accumulated, it should be possible to test our model in the future. 

We should note that in our discussion $J_0$ and $\gamma$ are just adjusted visually to be consistent with the low-energy data points. Dedicated investigation would require global fitting, which is beyond our scope here. Just for comparison, IceCube~\cite{Aartsen:2015knd} gives the best-fit parameters of a single unbroken power law for neutrinos energies between 25TeV and 2.8PeV without DM contribution, $J_0=6.7^{+1.1}_{-1.2}\times 10^{-8}\GeV$/cm$^2$/s/sr and $\gamma=0.5\pm 0.09$. 

\begin{figure}[t]
\includegraphics[width=0.40\textwidth, height=0.41\textwidth]{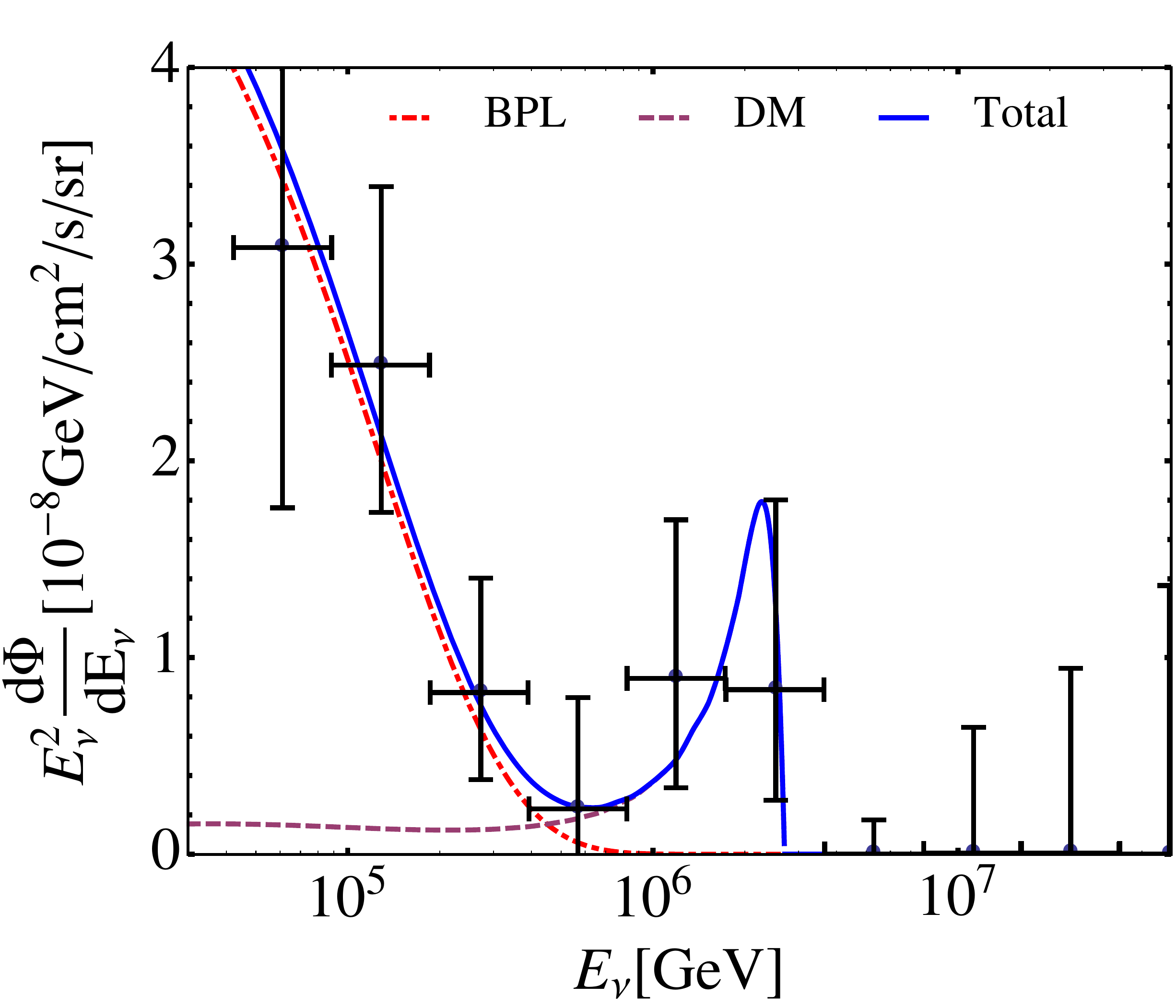} 
\includegraphics[width=0.40\textwidth, height=0.41\textwidth]{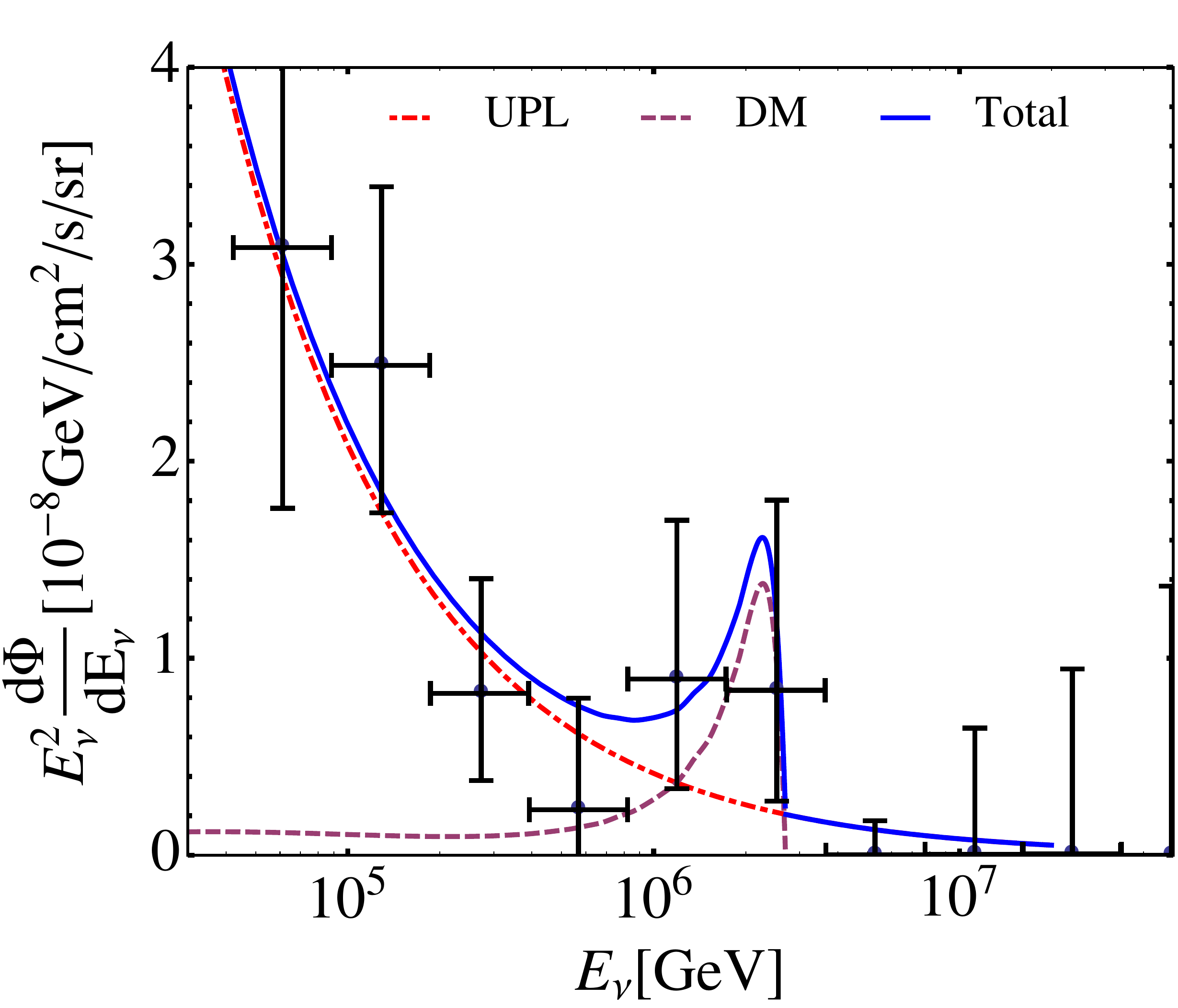}
\caption{Same as Fig.~\ref{fig:flux} but with preliminary updated results based on 4-year data~\cite{4IceCube}, with $J_0^\mathrm{{BPL}}=5.6\times 10^{-8}\GeV$/cm$^2$/s/sr and $\tau_\chi \sim 1.5\times 10^{28}$s (left) , and $J_0^\mathrm{{UPL}}=2.1\times 10^{-8}\GeV$/cm$^2$/s/sr and $\tau_\chi \sim 2\times 10^{28}$s (right). \label{fig:flux2}}
\end{figure}

\begin{figure}[t]
\includegraphics[width=0.40\textwidth, height=0.41\textwidth]{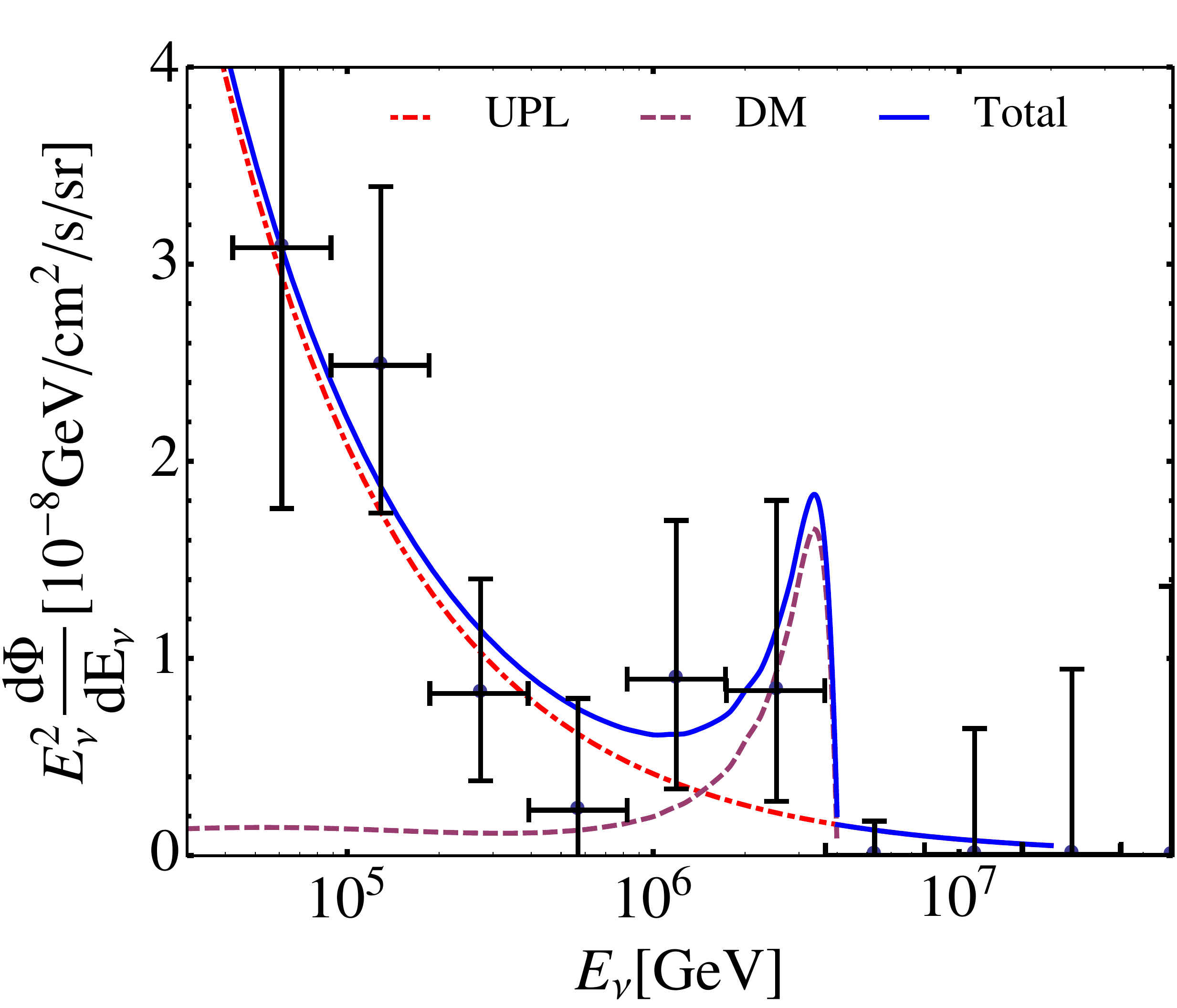} 
\includegraphics[width=0.40\textwidth, height=0.41\textwidth]{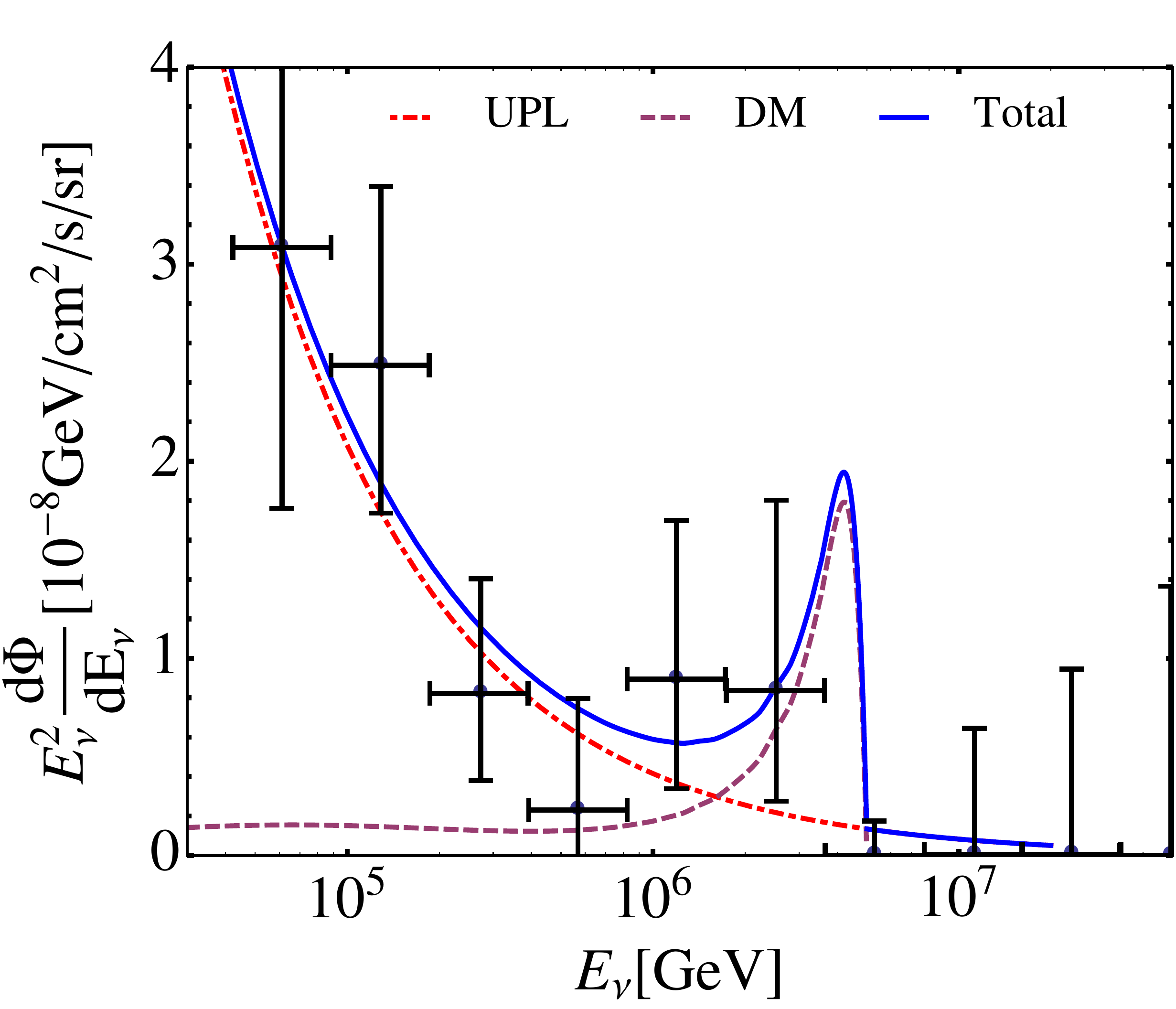}
\caption{Same as the right panel of Fig.~\ref{fig:flux2} but with $m_\chi=8\PeV, \tau_\chi \sim 1.7\times 10^{28}$s (left), and $m_\chi=10\PeV, \tau_\chi \sim 1.5\times 10^{28}$s (right). \label{fig:flux3}}.
\end{figure}

In Fig.~\ref{fig:flux2}, we also compare our model with the preliminary updated results~\cite{4IceCube} based on IceCube 4-year data which has already filled the gap a bit. Here we have only shifted to $J_0^\mathrm{{BPL}}=5.6\times 10^{-8}\GeV$/cm$^2$/s/sr and $\tau_\chi \sim 1.5\times 10^{28}$s (left panel), and  $J_0^\mathrm{{UPL}}=2.1\times 10^{-8}\GeV$/cm$^2$/s/sr and $\tau_\chi \sim 2\times 10^{28}$s (right panel). In Fig.~\ref{fig:flux3}, we illustrate two cases, one with $m_\chi=8\PeV, \tau_\chi \sim 1.7\times 10^{28}$s, and the other $m_\chi=10\PeV, \tau_\chi \sim 1.5\times 10^{28}$s.

There are some crucial differences between our model and some others in the literature. 
For example, the authors in Ref.~\cite{Yanagida:2013kka, Higaki:2014dwa} considered the effective operator, 
$y\bar{L}\widetilde{H}\chi$ with $y\sim 10^{-30}$, which induces mainly two-body decay of DM $\chi$, 
\[
\chi\rightarrow \nu h,\; \nu Z,\; l^{\pm}W^{\mu}.
\]
In this scenario, the neutrino spectrum shows that there should be no gap between 
$400~\TeV \sim 1~\PeV$~\cite{Esmaili:2014rma}.
Our model predicts that the dominant decay mode are
\[
\chi \rightarrow \phi/Z' + h +\nu,\; \phi/Z' + Z + \nu,\; \phi/Z' + W^{\pm} +l^{\mp},
\]
which is a consequence of $U(1)_X$ dark gauge symmetry and the dark charge 
assignments  of the dark Higgs and dark matter fermion $\chi$.  The neutrino spectra  
from primary $\chi$ decay and the secondary decays of $h$ and $\phi$ have different 
shapes and could account for the possible gap. However, we should note that the current 
data can not favor one over another yet due to its low statistics.  
Also the neutrino flux in our model is softer than the one predicted in  Ref.~\cite{Yanagida:2013kka, Higaki:2014dwa}, for example.

In Ref.~\cite{Boucenna:2015tra}, leptophilic three-body decay induced by dimension-six 
$\bar{L}_\alpha l_\beta \bar{L}_\gamma \chi$ was considered with global $U(1)$ or $A_4$ 
flavor symmetries. Besides the neutrino spectrum difference, our model involves an additional gauge boson which mediates the DM-nucleon scattering, and could be tested by DM direct searches.

Our scenario is also different from those in which DM decay is also responsible for the low-energy flux~\cite{Esmaili:2013gha}. The DM lifetime in Ref.~\cite{Esmaili:2013gha} should be around $2\times 10^{27}$s, as mainly determined by the low energy part of events. This is partly due to the reason that the branching ratio into neutrinos and $b\bar{b}$ there should be about $10 \%$ and $90 \%$, respectively, to account for the possible gap. On the other hand, in our scenario $1/2$ of the decay channels have prompt neutrinos. Another main difference is that three-body-decay usually gives broader spectra at PeV range than two-body-decay considered in Ref.~\cite{Esmaili:2013gha}, but more data is required in order to discriminate this difference.

Assuming the dark photon $X_\mu$ is much heavier than $Z$, the DM-nucleon scattering cross section 
can be roughly estimated as
\begin{equation}
\sigma_{\chi N}\sim \left(\frac{m^2_Z}{m^2_X}\right)^2\sin^2\epsilon\times 10^{-39}\textrm{cm}^2.
\end{equation}
$10^{-39}\textrm{cm}^2$ is the typical cross section value for SM $Z$-mediating DM-nucleon process. 
Comparing it with the direct detection bound for $100$GeV DM, we should have 
\begin{equation}\label{eq:ddbound}
\sigma_{\chi N} < 10^{-45}\textrm{cm}^2 \times \frac{m_\chi}{100 \GeV},
\end{equation}
for heavy $m_\chi \sim 5$ PeV.  This can be easily satisfied, for example, with $m_{X} \sim \TeV$ and $\sin \epsilon \lesssim 0.1$. 

%----------------------------------------------------------------------------------
\section{Relic Abundance and Constraints}\label{sec:relic}
%----------------------------------------------------------------------------------

In our above investigation, we have not discussed the relic abundance for DM $\chi$ yet. 
Since $\chi$ is very heavy, the unitary bound on its annihilation makes the DM $\chi$ impossible to 
be thermally produced and a non-thermal process is needed (see Ref.~\cite{Baer:2014eja} for a recent 
review on such topics). Here, we discuss one possible non-thermal production mechanism 
for DM $\chi$ in our model. We assume that the dark Higgs $\phi$~\footnote{It should be $\Phi$ precisely since 
at high temperature symmetries are not yet broken, but it will not affect our discussion.} once shared a common temperature with SM particles and had a thermal distribution when its temperature $T$ was larger than $m_\chi$. Here we do not specify the mechanism how $\phi$ reached such a temperature; it could be due to reheating after inflation or some heavy particle decays.  

In the thermal bath, $\chi$ could be produced through 
$\phi + \phi \rightarrow \chi +\bar{\chi},$  whose thermal cross section is given by
\[
\langle \sigma v\rangle \sim \frac{1}{16\pi}\left(\frac{f^2}{m_N}\right)^2. 
\]
Here we considered the $m_N\gg T$ case only. We can calculate the $\chi$'s yield, 
$Y_\chi\equiv n_{\chi+\bar{\chi}}/s$, where $s\sim g_{\ast s}(T)T^3$ is the total entropy 
density in the Universe ( $g_{\ast s}(T)\sim 100$),
\begin{equation}
Y_\chi \sim n_{\phi}\langle \sigma v\rangle /\mathcal{H}\sim \frac{T M_\mathrm{pl}\langle \sigma v\rangle }{g_{\ast s}(T)\sqrt{g_{\ast}(T)}}.
\end{equation}
In the above derivation we have used the Hubble parameter $\mathcal{H}\simeq  \sqrt{g_{\ast}(T)}T^2/M_\mathrm{pl}$, $g_{\ast}(T)\sim 100$ is the total effective number of degree of freedom when $\phi$'s temperature is $T$ and Planck mass $M_\mathrm{pl}=\sqrt{3/8\pi G}\simeq 4.2 \times 10^{18}~\GeV$. Since the yield has a positive power dependence on temperature, $\chi$ is mostly produced at $\phi$'s highest temperature $T^{\phi}_{\mathrm{max}}$,
\begin{equation}
Y_\chi \sim \langle \sigma v\rangle \times \frac{T^{\phi}_{\mathrm{max}} M_\mathrm{pl}}{g_{\ast s}(T^{\phi}_{\mathrm{max}})\sqrt{g_{\ast}(T^{\phi}_{\mathrm{max}})}}.
\end{equation}
Requiring  $\chi$ give the correct relic density, we have a relation
\begin{equation}
Y_\chi \sim 6\times 10^{-10} \left(\frac{\Omega_{\chi}}{\Omega_{\textrm{b}}}\right) \left(\frac{\GeV}{m_\chi}\right)\sim 3\times 10^{-15}\left(\frac{\PeV}{m_\chi}\right),
\end{equation}
which  puts a constraint on $f$ and $m_N$,
\begin{equation}\label{eq:fnrelation}
\left(\frac{f^2}{m_N}\right)^2\simeq 1.5\times 10^{-13}\left(\frac{\PeV}{m_\chi}\right)\times \frac{g_{\ast s}(T^{\phi}_{\mathrm{max}}) \sqrt{g_{\ast}(T^{\phi}_{\mathrm{max}})}}{T^{\phi}_{\mathrm{max}} M_\mathrm{pl}}. 
\end{equation}
When $m_N>T^{\phi}_{\mathrm{max}}>m_\chi$ and $m_\chi\sim \PeV$, we are able to give a lower bound for $f$ 
\begin{equation*}
|f|\gtrsim 10^{-6}.
\end{equation*}

One more thing we can infer from Eq.~(\ref{eq:yfnrelation}) and (\ref{eq:fnrelation}) is that in this production mechanism $y$ would be too small such that this right-hand neutrino $N$ can not be fully responsible for active neutrino mass and mixing angles. This is because, in order to explain the active neutrino mass, we should have 
\begin{equation}
\frac{y^2v^2_H}{m_N}\simeq 0.1 \mathrm{eV}\Rightarrow y \sim 10^{-5} \left(\frac{m_N}{\PeV}\right)^{1/2},
\end{equation}
which can not be satisfied simultaneously with Eq.~(\ref{eq:yfnrelation}) and (\ref{eq:fnrelation}). 
This is not a problem for this model since we can expect there are additional right-handed neutrinos $N_i$ 
and they can couple to $\bar{L}\widetilde{H}$ with large $y_i$ but to $\bar{\chi}\Phi$ with tiny $f_i$, 
so that they are just responsible for active neutrino mass and mixing angle but not for DM $\chi$'s production.

DM direct detection can constrain DM-nucleon scattering cross section, whose value in our model is determined by the kinetic mixing parameter $\epsilon$, gauge coupling $g_X$ 
and the mass of $Z'$. For GeV-TeV DM, there is already plenty of viable parameter space 
to evade such a constraint, see Ref.~\cite{Ko:2014nha} for example. For PeV DM, the 
constraint is even relaxed due to the low number density, see Eq.~\ref{eq:ddbound}. Indirect detection from positron, anti-proton and $\gamma$ rays also constrain DM $\chi$'s decay lifetime, see Refs.~\cite{Esmaili:2012us, Murase:2012xs, Ibarra:2013cra} and references therein for example. We have checked that $\tau_\chi\sim 10^{28}$ sec is still allowed by all such constraints. 

\begin{figure}[t]
\includegraphics[width=0.60\textwidth, height=0.5\textwidth]{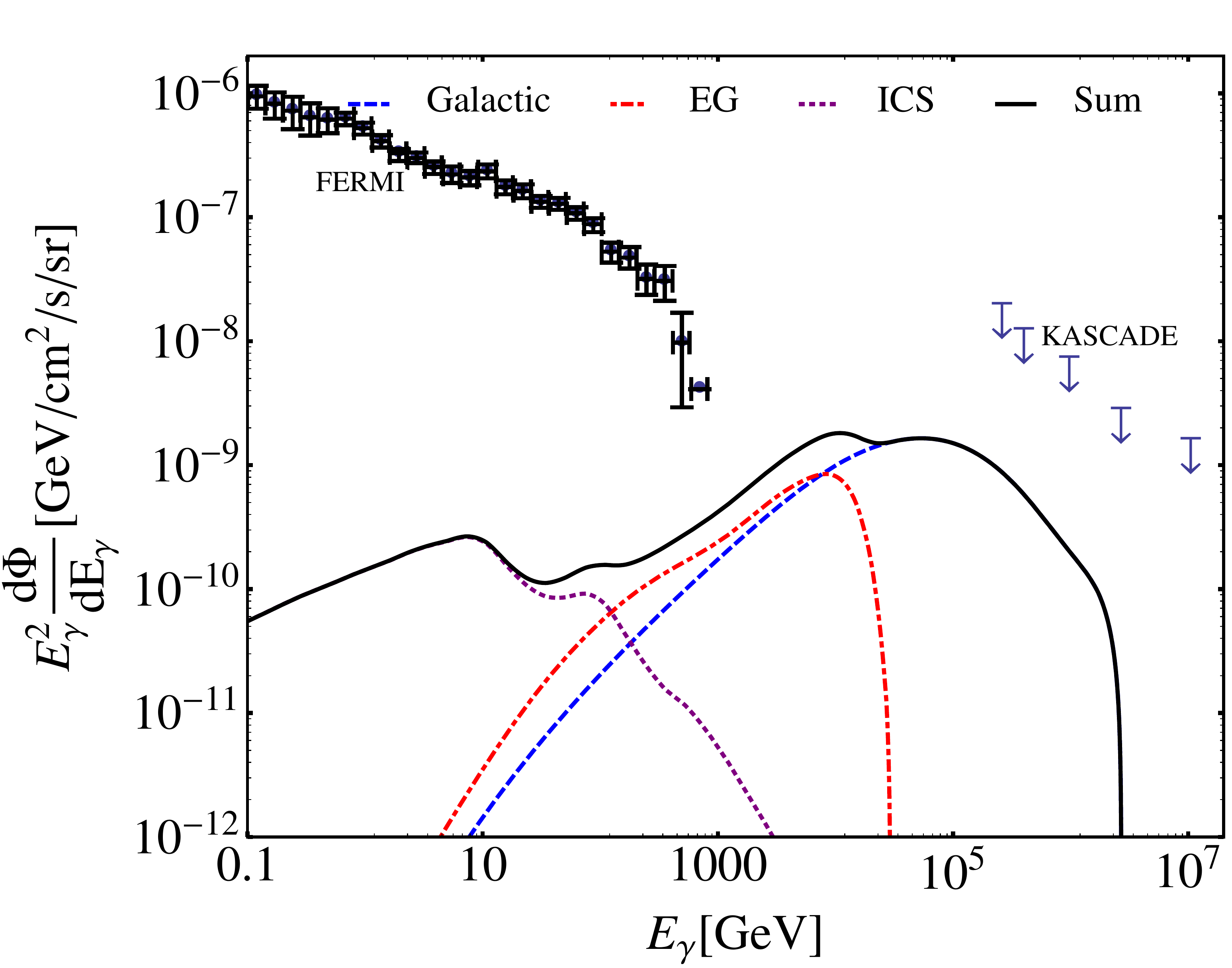} 
\caption{The gamma-ray flux from DM decay with $m_\chi\sim 5\PeV$ and lifetime $\tau_\chi \sim 2\times 10^{28}$s, confronted with constraints from Fermi-LAT~\cite{FERMI2014:IGRB} and KASCADE~\cite{KASCADE:2003} data. \label{fig:gamma}}
\end{figure}

As an illustration, in Fig.~\ref{fig:gamma} we show the expected gamma-ray flux from DM decay with $m_\chi\sim 5\PeV$ and lifetime $\tau_\chi \sim 2\times 10^{28}$s. We have included the prompt gamma-rays and those from inverse compton scattering (ICS) of charged particles on CMB, starlight and dust-rescattered light. For extragalatic contribution, we have taken absorption factor into account. As we can easily see, the flux is well below the current constraints from Fermi-LAT~\cite{FERMI2014:IGRB} and KASCADE~\cite{KASCADE:2003}. Our results are also consistent with the recent investigations about gamma-ray constraints on the lifetime of PeV DM in different scenarios~\cite{Murase:2015gea,Esmaili:2015xpa}, $\tau_\chi \gtrsim 3\times 10^{27}$.

%----------------------------------------------------------------------------------
\section{Conclusion}\label{sec:conclusion}
%----------------------------------------------------------------------------------
In this paper, we have proposed a dark matter (DM) model that can explain the IceCube 
PeV events in terms of DM decay.  The model is based on an extra $U(1)_X$ dark gauge 
symmetry which is spontaneously broken  by a dark Higgs field, $\Phi$. 
One crucial bridge between DM $\chi$ and standard model (SM) particles is established 
by heavy right-handed (RH) neutrino portal interactions.  This heavy neutrino can induce 
DM decays into SM particles, including the light neutrinos. 

The dominant decay channel of DM is the three-body final state with 
%decaying process where DM particle decays into 
SM Higgs, dark Higgs, and neutrino ($\chi\rightarrow \phi + h +\nu$)
(and other channels due to the Goldstone boson equivalence theorem), and not the usual two-body decays such as $\chi \rightarrow Z \nu, W^\pm l^\mp$, etc.. 
This is a unique feature of the present model based on $U(1)_X$ dark gauge symmetry 
and the RH neutrino portal interactions. We have calculated both total and differential 
decay width to evaluate the galactic and extragalactic neutrino fluxes. 
We have found that neutrino flux from these decay products can agree well with the 
IceCube spectrum. Together with an astrophysical flux for lower energy events, 
we are able to fit IceCube data around $100~\TeV\sim 2~\PeV$ if we assume DM mass is about $ m_\chi \sim 5$PeV and its lifetime is $\tau_\chi \sim 2\times 10^{28}$ sec.

\begin{acknowledgments} 
PK would like to thank Jonathan L. Rosner and Thomas J. Weiler for carefully reading the manuscript and making numerous comments, and Celine Boehm, Aelxander Kusenko and Steven Parke for useful discussions on the IceCube events. 
Y.~Tang would like to thank Eung-Jin Chun, Jia Liu and Xiao-Ping Wang for helpful discussions, and express a special thanks to Mainz Institute for Theoretical Physics for its hospitality and support. 
This work is supported in part by National Research Foundation of Korea (NRF) Research 
Grant NRF-2015R1A2A1A05001869, by the NRF grant funded by the Korea government 
(MSIP) (No. 2009-0083526) through Korea Neutrino Research Center at Seoul National 
University, and by National Science Foundation Grant No. PHYS-1066293 and 
the hospitality of the Aspen Center for Physics where this work has been completed (PK).
\end{acknowledgments}

\section*{Appendix}
Here we show the complete differential decay width for $\chi\rightarrow h + \phi + \nu$. Throughout the calculation, we work in the rest frame of $\chi$, so $\chi$'s momentum is $(m_\chi,0,0,0)$. For unpolarized $\chi$, we have
%\begin{widetext}
\begin{align}
d\Gamma=&\frac{1}{\left(2\pi\right)^3}\frac{1}{8m_\chi}\sum_{pol} \overline{|\mathcal{M}|^2} dE_\nu dE_h=\frac{1}{\left(2\pi\right)^3}\left(\frac{yf}{2m_N}\right)^2 E_\nu dE_\nu dE_h,
\end{align}
%\end{widetext}
where we have used the averaged, squared matrix element,
\[
\sum_{pol}|\mathcal{M}|^2 = 2\left(\frac{yf}{m_N}\right)^2 \left(m_\chi E_\nu -m^2_\nu\right)\simeq 2\left(\frac{yf}{m_N}\right)^2 m_\chi E_\nu.
\]
Then we get
%\begin{widetext}
\begin{align}
\frac{d\Gamma}{dE_\nu}=&\frac{E_\nu }{\left(2\pi\right)^3}\left(\frac{yf}{2m_N}\right)^2 \int _{E_h^{\mathrm{min}}} ^{E_h^{\mathrm{max}}}dE_h \nonumber  \\
=&\frac{E_\nu }{\left(2\pi\right)^3} \left(\frac{yf}{2m_N}\right)^2 \frac{\left(m^2_{\phi\nu}\right)_{\mathrm{max}}-\left(m^2_{\phi \nu}\right)_{\mathrm{min}}}{2m_\chi},\nonumber\\
=&\frac{2E_\nu }{\left(2\pi\right)^3m_\chi} \left(\frac{yf}{2m_N}\right)^2 \sqrt{E^{\ast 2}_\phi - m^2_\phi}\sqrt{E^{\ast 2}_\nu - m^2_\nu},\\
\simeq &\frac{E^2_\nu }{\left(2\pi\right)^3} \left(\frac{yf}{2m_N}\right)^2  
\left( {\rm with} ~E_\nu<m_\chi/2 \right)  .
\end{align}
%\end{widetext}
In the last line we have used the heavy $m_\chi$ limit, $m_\chi \gg m_h , m_\chi \gg m_{phi}$ and $m_\nu \simeq 0$. Some definitions are listed below,
\begin{eqnarray*}
E_h=\frac{m^2_\chi + m^2_h - m^2_{\phi \nu}}{2m_\chi}, 
\\ m^2_{ab}=\left(p_a+p_b\right)^2, 
m_{ab}=\sqrt{\left(p_a+p_b\right)^2},
\end{eqnarray*}
and some other kinematic variables,
\begin{align*}
\left(m^2_{\phi \nu}\right)_{\mathrm{max}} = & \left(E^{\ast}_\phi + E^{\ast}_\nu\right)^2 -
\left(\sqrt{E^{\ast 2}_\phi - m^2_\phi}-\sqrt{E^{\ast 2}_\nu - m^2_\nu}\right)^2,\\
\left(m^2_{\phi \nu}\right)_{\mathrm{min}} = & \left(E^{\ast}_\phi + E^{\ast}_\nu\right)^2 - 
\left(\sqrt{E^{\ast 2}_\phi - m^2_\phi}+\sqrt{E^{\ast 2}_\nu - m^2_\nu}\right)^2,\\
E^{\ast}_\phi =& \frac{m^2_{h\phi}-m^2_h+m^2_\phi}{2m_{h\phi}} ,
E^{\ast}_\nu = \frac{m^2_\chi-m^2_{h\phi}+m^2_\nu}{2m_{h\phi}} ,\\
m^2_{h\phi}=& m^2_\chi + m^2_\nu - 2 m_\chi E_\nu,
\end{align*}
where $E^{\ast}_\phi$ and $E^{\ast}_\nu$ are the energies of $\phi$ and $\nu$ in the $m_{h\phi}$ rest frame, respectively. From the above differential decay width, we can easily get the total width
\begin{equation}
\Gamma\simeq \frac{1}{24}\frac{m^3_\chi}{\left(2\pi\right)^3} \left(\frac{yf}{2m_N}\right)^2.
\end{equation}

The differential decay width as function of $E_h$ or $E_\phi$ can also be calculated similarly. For example,
\begin{align}
\frac{d\Gamma}{dE_h}=&\frac{1}{\left(2\pi\right)^3}\left(\frac{yf}{2m_N}\right)^2 \int _{E_\nu^{\mathrm{min}}} ^{E_\nu^{\mathrm{max}}}E_\nu dE_\nu \nonumber \\
=& \frac{1}{2\left(2\pi\right)^3} \left(\frac{yf}{2m_N}\right)^2 \left[ \left(E^{\mathrm{max}}_{\nu}\right)^2 - \left(E^{\mathrm{min}}_{\nu}\right)^2\right],\\
\simeq &\frac{ E_h \left(m_\chi - E_h\right)}{2 \left(2\pi\right)^3}\left(\frac{yf}{2m_N}\right)^2, E_h<m_\chi/2.
\end{align}
Again, in the second line we have used the massless limit. The neutrino energy $E_\nu$ in $\chi$'s rest frame can be written as
\begin{eqnarray}
E_\nu=\frac{m^2_\chi + m^2_\nu - m^2_{\phi h}}{2m_\chi},
\end{eqnarray}
with similar definitions and kinematic bounds,
\begin{align*}
\left(m^2_{\phi h}\right)_{\mathrm{max}} = & \left(E^{\ast}_\phi + E^{\ast}_h\right)^2 -
\left(\sqrt{E^{\ast 2}_\phi - m^2_\phi}-\sqrt{E^{\ast 2}_h - m^2_h}\right)^2,\\
\left(m^2_{\phi h}\right)_{\mathrm{min}} = & \left(E^{\ast}_\phi + E^{\ast}_h\right)^2 - 
\left(\sqrt{E^{\ast 2}_\phi - m^2_\phi}+\sqrt{E^{\ast 2}_h - m^2_h}\right)^2,\\
E^{\ast}_\phi =& \frac{m^2_{\phi\nu}-m^2_\phi+m^2_\nu}{2m_{\phi\nu}} ,
E^{\ast}_h = \frac{m^2_\chi-m^2_{\phi\nu}+m^2_\phi}{2m_{\phi\nu}} ,\\
m^2_{\phi\nu}=& m^2_\chi + m^2_h - 2 m_\chi E_h,
\end{align*}
now here $E^{\ast}_\phi$ and $E^{\ast}_h$ are the energies of $\phi$ and $h$ in the $m_{\phi \nu}$ rest frame, respectively.

%\bibliographystyle{abbrv}
%\bibliographystyle{../utphysMa}
%\bibliography{../references}

\providecommand{\href}[2]{#2}\begingroup\raggedright\endgroup

\end{document}